\documentclass[%
 reprint,
superscriptaddress,
 amsmath,amssymb,
 aps,
]{revtex4-2}

\usepackage{graphicx}
\usepackage{dcolumn}
\usepackage{bm}
\usepackage{hyperref}
\hypersetup{
    colorlinks=true
}

\usepackage{braket}

\usepackage{algorithm}
\usepackage{algorithmicx}
\usepackage{algpseudocode}
\usepackage{graphicx}

\algnewcommand{\Initialize}[1]{%
  \State \textbf{Initialize:}
  \State \hspace*{\algorithmicindent}\parbox[t]{0.8\linewidth}{\raggedright #1}
}

\algdef{SE}[DOWHILE]{Do}{doWhile}{\algorithmicdo}[1]{\algorithmicwhile\ #1}%

\newcommand{\norm}[1]{\|{#1}\|}
\newcommand{\abs}[1]{|{#1}|}
\newcommand{\I}[0]{\mathrm{I}}
\newcommand{\II}[0]{\mathrm{I}\hspace{-1.2pt}\mathrm{I}}

\newtheorem{thm}{Theorem}

\begin{document}

\title{Quantum Multi-Resolution Measurement with application to Quantum Linear Solver}

\author{Yoshiyuki Saito}
\altaffiliation[d8241104@u-aizu.ac.jp]{}
\affiliation{University of Aizu, Fukushima, Japan}

\author{Xinwei Lee}
\author{Dongsheng Cai}
\affiliation{University of Tsukuba, Ibaraki, Japan}

\author{Nobuyoshi Asai}
\affiliation{University of Aizu, Fukushima, Japan}

\date{\today}

\begin{abstract}
Quantum computation consists of a quantum state corresponding to a solution, and measurements with some observables.
To obtain a solution with an accuracy $\epsilon$, measurements $O(n/\epsilon^2)$ are required, where $n$ is the size of a problem.
The cost of these measurements requires a large computing time for an accurate solution.
In this paper, we propose a quantum multi-resolution measurement (QMRM), which is a hybrid quantum-classical algorithm that gives a solution with an accuracy $\epsilon$ in $O(n\log(1/\epsilon))$ measurements using a pair of functions.
The QMRM computational cost with an accuracy $\epsilon$ is smaller than $O(n/\epsilon^2)$.
We also propose an algorithm entitled QMRM-QLS (quantum linear solver) for solving a linear system of equations using the Harrow-Hassidim-Lloyd (HHL) algorithm as one of the examples.
We perform some numerical experiments that QMRM gives solutions to with an accuracy $\epsilon$ in $O(n\log(1/\epsilon))$ measurements.

\end{abstract}

\maketitle

\section{Introduction}
High-speed quantum computing is promising in many areas of the sciences and engineering in the near future because of its potential to solve complex tasks faster than classical computing. 
In fact, many quantum algorithms considered to be faster than conventional algorithms have been proposed, such as Shor's prime factoring algorithm \cite{shor1994Algorithmsb}, Grover's database search algorithm \cite{grover1996Fast}, the Hallow-Hassidim-Lloyd (HHL) algorithm solving a linear system of equations \cite{harrow2009Quantum}, etc.
In addition, many hybrid quantum algorithms have been applied to quantum chemical problems \cite{peruzzo2014Variationala, liu2019Variationala}, combinatorial optimization problems \cite{farhi2014Quantum, hadfield2019Quantum}, machine learning \cite{biamonte2017Quantum, schuld2015Introduction}, etc.
For some problems such as solving a linear system of equations and quantum chemical problems, high-accuracy computing is required in addition to high-speed computing.

It is essential to construct a highly accurate quantum state by accurate gate operations and measuring the quantum state until the required accuracy is obtained.
To obtain a solution with an accuracy $\epsilon$, measurements $O(1/\epsilon^2)$ are required \cite{mcclean2016Theorya}.
In the case of the problem size $n$, measurement $O(n/\epsilon^2)$ is needed.
In addition, the accuracy $\epsilon$ of the solution cannot exceed the accuracy of its quantum state.
This means $\norm{\ket{x} - \ket{\tilde{x}}} < \epsilon$, where $\norm{\cdot}$ denotes 2-norm, $\ket{x}$ is the exact quantum state, and $\ket{\tilde{x}}$ is the approximation state of $\ket{x}$ constructed on a quantum computer. 
In other words, the accuracy of the quantum state determines the upper limit of the solution accuracy.
In addition, to produce a highly accurate quantum state, many quantum resources such as quantum bits (qubits), gates, and circuit depths are required.
For practical quantum computing, the number of measurements is an important factor besides quantum resources. 
This is because computational time is proportional to the number of measurements.
Thus, it is necessary to reduce the number of measurements for high-accuracy computing.
In so doing, $O(n/\epsilon^2)$ measurements must be reduced corresponding to the accuracy $\epsilon$ or the number $n$.
Both quantum amplitude amplification and estimation methods have been proposed to reduce the samples of the problems using the Monte Carlo simulation \cite{montanaro2015Quantum, brassard2002Quantum, suzuki2020Amplitude, grinko2021Iterative}.
These achieve $O(1/\epsilon)$ samples to obtain a quadratic speedup \cite{brassard2002Quantum, suzuki2020Amplitude, grinko2021Iterative}.
Pauli operators grouping methods for simultaneous measurements can reduce the number of terms $n$ to be measured \cite{izmaylov2020Unitary, gokhale2020Measurementa, zhao2020Measurement}.
Although it is important to reduce the number of terms $n$, it is more important to reduce the number of measurements depending on the accuracy $\epsilon$, because the inverse of $\epsilon$ grows faster than $n$ for high-accuracy computation.

In this paper, we propose quantum multi-resolution measurement (QMRM) that give $O(n\log(1/\epsilon))$ measurements for a solution using $\epsilon$ accuracy employing a pair of functions.
The measurement cost of QMRM $O(n\log(1/\epsilon))$ is smaller than $O(n/\epsilon^2)$.
In addition, QMRM is a hybrid algorithm consisting of a quantum computation part and a classical computation part using a pair of functions.
In reality, QMRM is a generalization of \cite{saito2021Iterative}, and we focus on the number of measurements in the present paper.

For one of the applications of QMRM, we propose an algorithm entitled QMRM-QLS for solving a linear system of equations with a quantum algorithm, e.g., HHL.
Quantum algorithms for solving a linear system of equations, called Quantum Linear Solver(s) (QLS), are proposed \cite{harrow2009Quantum, ambainis2012Variable, childs2017Quantum, wossnig2018Quantum, bravo-prieto2020Variational, an2022Quantuma}.
To solve a linear system of equations $Ax = b$, full quantum algorithms \cite{harrow2009Quantum, ambainis2012Variable, childs2017Quantum, wossnig2018Quantum} compute $\ket{x} = A^{-1}\ket{b} / \norm{A^{-1}\ket{b}}$, where $A \in \mathbb{C}^{n \times n}$, $x, b \in \mathbb{C}^{n}$, and $\ket{x}, \ket{b}$ are, respectively, normalized vectors of $x$ and $b$.
The runtime cost of QLS roughly scales up to be $O(\log n)$ \cite{harrow2009Quantum, ambainis2012Variable, childs2017Quantum, wossnig2018Quantum}, which is exponentially faster than the conjugate gradient method ($O(n)$) \cite{shewchuk1994Introduction}.
However, since these quantum algorithms need many qubits and circuit depths, fault-tolerant quantum computers are desired for running these algorithms.
In addition to these full quantum algorithms, variational hybrid quantum-classical linear solvers that use parameterized quantum circuits adjusting their parameters are also proposed \cite{bravo-prieto2020Variational, an2022Quantuma}.

The present paper is organized as follows.
We introduce QMRM and show its measurement cost is $O(n\log(1/\epsilon))$ in Sec.~\ref{sec:framework}.
In Sec.~\ref{sec:review}, we briefly review classical and quantum algorithms for a linear system of equations.
Then, we propose QMRM-QLS for one of the applications of QMRM in Sec.~\ref{sec:applicaltion}.
In Sec.~\ref{sec:experiments}, we conduct numerical experiments to evaluate QMRM-QLS, the accuracy, and the number of measurements using $4 \times 4$ real Hermitian matrices.
Finally, we conclude QMRM and QMRM-QLS in Sec. \ref{sec:conclusion}.

\section{Quantum Multi-Resolution Measurement} \label{sec:framework}
In this section, first, we introduce quantum multi-resolution measurement (QMRM).
Next, we derive the costs of QMRM, the number of measurements for the solution with accuracy $\epsilon$, and quantum resources to run this quantum algorithm.

\subsection{The QMRM}
Quantum multi-resolution measurement (QMRM) is an iterative refinement method utilizing quantum computing.
QMRM employs a pair of functions that we call $f$ a problem generator and $g$ a solution adjustor in the iterative refinement process.
In QMRM, we solve a problem $f(x) = 0$, where $x$ is the solution.
When the $m$-th approximation $x_m$ of $x$ is given, QMRM refines $x_m$ to $x_{m+1}$ as follows:
\begin{align}
	x_{m+1} = x_m + g(\tilde{d}_m), \label{eq:solution_relation}
\end{align}
where $g(\tilde{d}_m)$ is an approximation of $d_m$ that is the exact solution of the problem $f(x_m) = 0$, which is solved in a quantum computer, and the adjustor $g$ is chosen to satisfy 
\begin{align} 
	\norm{x_{m+1} - x_{m}} < 1,
	\label{eq:convergence_condition_of_x}
\end{align}
so that $x_m$ converges to $x$.
The generator $f$ must satisfy the following condition:
\begin{align}
	\frac{\norm{f(x_{m+1})}}{\norm{f(x_{m})}} \leq \epsilon_m < 1, \label{eq:convergence_condition_of_f}
\end{align}
where $\epsilon_m$ is the accuracy in the $m$-th iterations.

Here, QMRM consists of two parts.
The classical part provides problems by the generator $f$ and gives the solution $x_{m+1}$ by the adjustor $g$ with measured values $\tilde{d}_m$, 
and a quantum part to solve problems using a quantum algorithm.
The QMRM procedure is summarized in Algorithm \ref{algo:main}.
The pair of $f$ and $g$ in the $m$-th iteration are, respectively, denoted by $f_m$ and $g_m$.
The inputs to QMRM are: 
\begin{itemize}
	\item[(1)] parameters to define $f_m$ and $g_m$, and to determine a quantum algorithm $\mathcal{Q}$ to solve the problem provided by $f_m$ in each step,
	\item[(2)] the target accuracy $\epsilon$ so that $\norm{f_m(x_{m})} \leq \epsilon < 1$,
	\item[(3)] the accuracy $\tilde{\epsilon}$ so that $\norm{\ket{d_m} - \ket{\tilde{d}_m}} \leq \tilde{\epsilon} < 1$, where $\ket{d_m}$ is the exact quantum state corresponding to the solution $d_m$ so that $x = x_m + d_m$ and $\ket{\tilde{d}_m}$ is an approximation of $\ket{d_m}$, and
	\item[(4)] the number of measurements $N_{\text{shots}}$ to obtain approximation $\tilde{d}_m$ from the quantum state $\ket{\tilde{d}_m}$.
\end{itemize}

Using these inputs, QMRM performs the following three tasks in each iteration:
\begin{itemize}
	\item[(a)] calculate $r_m = f_m(x_m)$,
	 \item[(b)] obtain the approximation $\tilde{d}_m$ by running the quantum algorithm $\mathcal{Q}$ and measuring $N_{\text{shots}}$ times, and
	 \item[(c)] calculate $g_m(\tilde{d}_m)$ to obtain the solution $x_{m+1}$.
\end{itemize}
Finally, QMRM gives the solution $x_m$ when $\norm{f_m(x_m)} \leq \epsilon$.

Here, the accuracy of $\tilde{d}_m$ with $N_{\text{shots}}$ is denoted by $\epsilon'$.
Since $\epsilon'$ depends on $N_{\text{shots}}$, its lower limit is the running accuracy $\tilde{\epsilon}$ of the quantum algorithm $\mathcal{Q}$.
The target accuracy $\epsilon$, the running accuracy $\tilde{\epsilon}$, the approximated accuracy $\epsilon'$ of the problem, and the accuracy in the $m$-th iteration $\epsilon_m$ have the following relation: 
\begin{align}
	\epsilon \ll \tilde{\epsilon} \leq \epsilon' \leq \epsilon_m. \label{eq:accuracy_relation}
\end{align}
Thus, $\mathcal{Q}$ is expressed as $\mathcal{Q}(\tilde{\epsilon}, p, r_m)$, where $p$ is the input parameter needed to solve the problem.

\begin{algorithm}[H]
	\caption{Quantum Multi-Resolution Measurement}
	\label{algo:main}
	\begin{algorithmic}[1]
		\Require{parameters $p$, target accuracy $\epsilon$, the number of measurements $N_{\text{shots}}$, running accuracy $\tilde{\epsilon}$}
		\Initialize{
			$x_0 = 0$ // initial guess

			$m = 0$ // iteration count

		}
		
		\While{True}
			\State (a) $r_m = f_m(x_{m})$
			
			\If{$\norm{r_m} < \epsilon$}
				\State break
			\EndIf		
		
			\State (b) {\it approximation $\tilde{d}_m$ by running $\mathcal{Q}(\tilde{\epsilon}, p, r_m)$ and measuring $N_{\text{shots}}$}
			
			\State (c) $x_{m+1} = x_m + g_m(\tilde{d}_m)$
		
			\State $m \leftarrow m + 1$
		
		\EndWhile

		\Ensure{solution $x$}
		
	\end{algorithmic}
\end{algorithm}

The generator $f$ and the adjustor $g$ can be characterized as follows.
The solution of the quantum solver is given by measurements.
The adjustor $g$ refines the current solution $x_m$ using the measured value $\tilde{d}_m$.
The generator $f$ provides a problem with $x_m$ to refine $x_m$ using the adjustor $g$, iteratively.
Here, we focus on two computing domains:
(i) a quantum computing domain that finds a solution using a quantum algorithm,
and (ii) a solution domain where a solution exists.
In other words, (i) the quantum computing domain is the Hilbert space on a quantum computer, and (ii) the solution domain is the computing domain to adjust and refine a solution.
The output $\tilde{d}_m$ belongs to the quantum computing domain (i) because it is obtained by measurements.
However, the solution domain (ii) is different from the quantum computing domain (i).
To obtain a solution from the output $\tilde{d}_m$ in the quantum computing domain (i), the adjuster $g$ transforms the output $\tilde{d}_m$ from the quantum computing domain (i) to the solution domain (ii).
The output $\tilde{d}_m$ has enough resolution elements which contribute to refining the current solution $x_m$ because $x_m$ is refined by $g(\tilde{d}_m)$.
Moreover, the generator $f$ defines a problem whose solution is close to the exact solution.
QMRM can perform measurements on various resolutions by iteratively adjusting the computing domain from (i) to (ii) using $f$ and $g$.

\subsection{Costs of QMRM}
The measurement cost of QMRM with an accuracy $\epsilon$ can be estimated by the following theorem:
\begin{thm} \label{theorem:measurements}
Given approximation $x_m$, a solution $\tilde{d}_m$ with $N_{\text{shots}} = O(n)$, a function $g$ satisfying Eqs.~(\ref{eq:solution_relation}) and (\ref{eq:convergence_condition_of_x}), and a function $f$ satisfying Eq.~(\ref{eq:convergence_condition_of_f}), the Algorithm \ref{algo:main} needs $N_\text{total} = O(n\log(1/\epsilon))$ measurements for the solution with accuracy $\epsilon$.
\end{thm}
Proof: The total number of measurements $N_\text{total}$ is written as $N_\text{shots} \times k$, where $k$ is the number of iterations needed to obtain the solution with accuracy $\epsilon$.
We require $k$ times the estimates.
After $k$ iterations, from Eq.~(\ref{eq:convergence_condition_of_f}), we have 
\begin{align}
	\norm{f_{k}(x_{k})} \leq \prod_{m=0}^{k-1} \epsilon_m \norm{f_0(x_0)}.
\end{align}
Here, we introduce $\epsilon_{\text{max}} = \max_{0\leq m \leq k-1} \epsilon_m$ and $C = \norm{f_0(x_0)}$.
We have
\begin{align}
	\norm{f_{k}(x_{k})} \leq C (\epsilon_{\text{max}})^{k} \leq \epsilon,
\end{align}
giving
\begin{align}
	k \geq \frac{\log(1/\epsilon)}{\log(1/\epsilon_{\text{max}})} - \frac{\log(1/C)}{\log(1/\epsilon_{\text{max}})}.
\end{align}
Since $\epsilon_{\text{max}}$ and $C$ are constant, we have $k = O(\log(1/\epsilon))$. $\blacksquare$

From theorem \ref{theorem:measurements}, the cost of QMRM including calculations $f$, $g$, and quantum algorithm $\mathcal{Q}$ is:
\begin{align}
	O(\log(1/\epsilon)) \times (O(n) + (\text{cost of }\mathcal{Q}) + (\text{cost of }f, g)) \nonumber \\
	= N_{\text{total}} + O(\log(1/\epsilon)) \times ((\text{cost of }\mathcal{Q}) + (\text{cost of }f, g)).
	 \label{eq:total_cost}
\end{align}

Since only finite $N_{\text{shots}}$ measurements are performed to obtain the solution, the running accuracy $\tilde{\epsilon}$ does not need to be set high enough compared with the accuracy $\epsilon'$.
Therefore, it is possible to run $\mathcal{Q}(\epsilon', \cdots)$ instead of $\mathcal{Q}(\tilde{\epsilon}, \cdots)$.
Thus, Eq.~(\ref{eq:accuracy_relation}) can be rewritten as
\begin{align}
	\epsilon \ll \tilde{\epsilon} \approx \epsilon' \leq \epsilon_m. \label{eq:accuracy_relation_v2}
\end{align}

The quantum circuit $\mathcal{Q}$ does not need to be constructed at each iteration.
This is because the quantum circuit consists of two components:
a fixed component using fixed parameters in all iterations and a varying component using variables such as $r_m$ per iteration.
Thus, QMRM reuses the fixed component, otherwise, it is reconstructed every iteration.

\section{Preliminary} \label{sec:review}
In this section, we introduce the iterative refinement method for a linear system of equations \cite{wilkinson1994Roundinga, moler1967Iterative} and the HHL quantum linear solver \cite{harrow2009Quantum} for QMRM-QLS that is described in the next section \ref{sec:applicaltion}.

\subsection{Iterative Refinement Method}
Here, we briefly introduce the iterative refinement method for a linear system of equations \cite{wilkinson1994Roundinga, moler1967Iterative}.
This iterative method is a practical technique to refine the computed solution $x_m$ that is an approximation of the exact solution $x$ satisfying a linear system $Ax = b$, where $A$ is a matrix and $b$ is a vector.
To refine the current solution $x_m$, first, calculate the residual as follows: 
\begin{align}
	r_m &= b - Ax_m. \label{eq:residual}
\end{align}
Then, solve the linear system:
\begin{align}
	A d_m = r_m, \label{eq:Ad=r}
\end{align}
and obtain the solution $d_m$.
Finally, update the solution as follows:
\begin{align}
	x_{m+1} = x_m + d_m. \label{eq:xm+1}
\end{align}

The residual Eq.(\ref{eq:residual}) requires $O(n^2)$ computational time, where $n$ is the size of the system.
If a matrix $A$ is sparse, $O(n)$ computational time is required to calculate the residual.
Some techniques use single-precision arithmetic to solve the system and use double-precision arithmetic to calculate residuals and to update the solution \cite{buttari2007Mixed, buttari2008Using}.
These techniques take advantage of single-precision arithmetic speed to solve the problem.

\subsection{The HHL algorithm}
Here, we introduce the HHL algorithm briefly \cite{harrow2009Quantum}.
To find a solution $x$ of a linear system of equations $Ax = b$, the HHL algorithm finds a quantum state $\ket{x}$ corresponding to $x$, where a matrix $A \in \mathbb{C}^{n\times n}$ is Hermitian and vectors $x, b \in \mathbb{C}^n$.
In the non-Hermitian case, a matrix $\tilde{A}$, we use:
\begin{align}
	\tilde{A} := 
	\begin{bmatrix}
		0   & A \\
		A^* & 0	
	\end{bmatrix}, 
\end{align}
which is always Hermitian, and we solve:
\begin{align}
	\tilde{A}\tilde{x} = \tilde{b}, 
\end{align}
using the HHL algorithm, where 
$\tilde{x} = 
\begin{bmatrix}
	0 & x^T 	
\end{bmatrix}^T$ and
$\tilde{b} =
\begin{bmatrix}
	b^T & 0
\end{bmatrix}^T
$.

The algorithm begins preparing a quantum state $\ket{\Psi} = \ket{b}\ket{0}_p \ket{0}_a$, where $\ket{b}$ is the unit vector of $b$, $\ket{0}_p$ is $p$ qubits register for the quantum phase estimation algorithm, and $\ket{0}_a$ is ancilla qubit.
Quantum phase estimation is performed with the unitary matrix $e^{iAt}$ as follows: 
\begin{align} \label{eq:after_qpe}
	\ket{\Psi} \mapsto \sum_{j=0}^{n-1} \beta_j\ket{u_j}\ket{\tilde{\lambda}_j}_p \ket{0}_a,
\end{align}
where parameter $t$ is chosen to map all eigenvalues of $A$, $\lambda_j$, to the interval $[0, 2\pi)$ to maintain a one-to-one correspondence from $\lambda_j$ to $e^{i\lambda_j t}$ of $e^{iAt}$.
Here, $\ket{u_j}$ is the eigenvector corresponding to the eigenvalue $e^{i\lambda_j t}$, and $\beta_j$ is the coefficient so that $\ket{b} = \sum_{j=0}^{n-1} \beta_j \ket{u_j}$.
Moreover, $\tilde{\lambda}_j$ is $\lambda_j$ scaled by constant $t$ and expressed by $p$ qubits register.
Then, apply an eigenvalue rotation to ancilla qubit:
\begin{align} \label{eq:rotation}
	\sum_{j=0}^{n-1} \beta_j\ket{u_j}\ket{\tilde{\lambda}_j}_p 
	\left( \sqrt{1 - \frac{C^2}{\tilde{\lambda}_j^2}}\ket{0}_a + \frac{C}{\tilde{\lambda}_j}\ket{1}_a \right),
\end{align}
where $C$ is a normalization constant.
After the rotation, the register $p$ is uncomputed by inverse phase estimation.
Then, the ancilla qubit $a$ is measured.
If the measurement is $\ket{1}_a$, the solution state $\ket{x}$ is obtained
\begin{align}
	C' \sum_{j=0}^{n-1} \frac{\beta_j}{\tilde{\lambda}_j}\ket{u_j} \equiv \ket{x}, \label{eq:x_HHL}
\end{align}
where $C'$ is a normalization constant.
If $\ket{0}_a$ is obtained, the HHL process failed and must be restarted.

Note that the accuracy of $\ket{x}$ is mined by the accuracy of the eigenvalue $\lambda_j$ that is determined as $1/2^m$ with $m$ qubits for the phase estimation in Eq.~(\ref{eq:after_qpe}) \cite[section 5.2]{nielsen2010Quantum}.
In addition, the quantum linear solver is useful to estimate expectation values $\braket{x|M|x}$, where $M$ are observables.
For a practical estimation, a full solution vector is essential for solving a linear system of equations.
To obtain a full solution vector $\ket{x}$ with accuracy $\epsilon$, we can use quantum state tomography at the cost $O(n(\log n)^2/\epsilon^2)$ \cite{gross2010Quantum}.

\section{QMRM Quantum Linear Solvers} \label{sec:applicaltion}
We propose an algorithm (QMRM-QLS) for solving a linear system of equations using the HHL algorithm.
In the following, we propose two types of QMRM-QLS using a different $f$ and $g$.  

\subsection{QMRM-QLS $\I$}
We propose a quantum linear solver entitled QMRM-QLS $\I$ summarized in Algorithm \ref{algo:IIMHHL} using QMRM (Algorithm \ref{algo:main}).
Here, QLS that computes $\ket{x}$ satisfying $A\ket{x} \propto \ket{b}$ with an accuracy less than $\tilde{\epsilon}$ is denoted by $\mathcal{QLS}(\tilde{\epsilon}, A, \ket{b})$.
The algorithm QMRM-QLS $\I$ is an iterative refinement method using $\mathcal{QLS}(\tilde{\epsilon}, A, \ket{b})$ as a linear solver engine.

Here, $f_m$ and $g_m$ are defined as follows:
\begin{align}
	f_m(x_m) &:= b - Ax_m, \label{eq:IIMHHL:f} \\
	g_m(\tilde{d}_m) &:= c_1 e^{ic_2} \tilde{d}_m, \label{eq:IIMHHL:g}
\end{align}
where $c_1 = \frac{\norm{r_m}}{\norm{A\tilde{d}_m}}$ that adjusts the scale of solution $x$ and $c_2 = \arg(r_m \cdot A\tilde{d}_m)$ adjusts the sign of vector $\tilde{d}_m$ as $e^{ic_2}$, where $r_m = f_m(x_m)$ is the residual vector and the operator `` $\cdot$ '' denotes the inner product.
To find the solution $x_m$ satisfying $Ax_m = r_m$, we use $\mathcal{QLS}(\tilde{\epsilon}, A, \ket{r_m})$ and perform $N_{\text{shots}}$ measurements for a solution state $\ket{\tilde{d}_m}$ by $Z$-basis measurements.
These measurements are required for an approximation $\tilde{d}_m = \sum_{j=0}^{n-1}\abs{\gamma_j}\ket{j}$, where $\gamma_j$ is the coefficient.
The adjusted value $g_m(\tilde{d}_m)$ is added to the current solution $x_m$.
These iterations are continued until the residual norm $\norm{r_m}$ converges.

From Eqs.~(\ref{eq:convergence_condition_of_x}) and (\ref{eq:xm+1}), the current solution $x_m$ can be refined when $\norm{d_m - g_m(\tilde{d}_m)} < 1$.
However, $x_m$ can not be refined when $\norm{d_m - g_m(\tilde{d}_m)} \geq 1$.
This case may occur because each element of $\tilde{d}_m$ has variances due to measurements.
Even if this case $\norm{d_m - g_m(\tilde{d}_m)} \geq 1$ occurs, the solution can be refined in the next iteration.

\begin{algorithm}[H]
	\caption{QMRM-QLS $\I$}
	\label{algo:IIMHHL}
	\begin{algorithmic}[1]

		\Require{matrix $A$, vector $b$, target accuracy $\epsilon$, the number of measurements $N_{\text{shots}}$, running accuracy $\tilde{\epsilon}$}
		\Initialize{
			$x = 0$ // initial guess

			$m = 0$ // iteration count

		}
		
		\While{True}
			\State (a) $r_m = f_m(x_m)$
			
			\If{$\norm{r_m} < \epsilon$}
				\State break
			\EndIf		
		
			\State (b) {\it approximation $\tilde{d}_m$ by running $\mathcal{QLS}(\tilde{\epsilon}, A, \ket{r_m})$ and measuring $N_{\text{shots}}$}
			
			\State (c) $x_{m+1} = x_m + g_m(\tilde{d}_m)$
		
			\State $m \leftarrow m + 1$
		
		\EndWhile

		\Ensure{solution $x$}

	\end{algorithmic}
\end{algorithm}

The quantum circuit of $\mathcal{QLS}(\tilde{\epsilon}, A, \ket{r_m})$, which is the HHL algorithm used for numerical experiments, consists of two components:
(1) the fixed component of the phase estimation Eq.~(\ref{eq:after_qpe}) and the eigenvalue rotation Eq.(\ref{eq:rotation}) using the matrix $A$,
and (2) varying components preparing the initial quantum state $\ket{r_m}\ket{0}_p\ket{0}_a$.
Thus, we only need to construct the variable part.

Since the accuracy is improved through iterative calculations, the norms of both the residual vector and the solution shrink.
However, the norm of the output $\tilde{d}_m$ is $1$.
Thus, $\tilde{d}_m$ must be adjusted using the constant $c_1$.
In the solution, both positive and negative elements are mixed.
Since the QMRM-QLS $\I$ uses only $Z$-basis measurements, the elements of the solution $\tilde{d}_m$ are all non-negative.
Thus, errors occur. 
To avoid the process being broken down due to the accumulation of errors, the sign of $\tilde{d}_m$ can be adjusted using $e^{ic_2}$.

Since the pair of $f_m$ and $g_m$, respectively, Eqs.~(\ref{eq:IIMHHL:f}) and (\ref{eq:IIMHHL:g}), perform mainly matrix-vector multiplication, their calculation costs are $O(n)$ for a sparse matrix and $O(n^2)$ for a dense matrix.
From Eq.~(\ref{eq:total_cost}), the cost of QMRM-QLS $\I$ for a sparse matrix is: 
\begin{align}
	O(\log(1/\epsilon)) \times (O(n)+\text{cost of }\mathcal{QLS} + O(n)). \label{eq:IIMHHL:cost}
\end{align}
For a sparse matrix, since the cost of the HHL algorithm is roughly $O(\log n)$, the cost Eq.~(\ref{eq:IIMHHL:cost}) is:
\begin{align}
	O(\log(1/\epsilon)) \times (O(n)+O(\log n) + O(n)) = O(n\log(1/\epsilon)) \label{eq:QMRMQLS_cost_sparse}
\end{align}
For a dense matrix, the cost of QLS \cite{wossnig2018Quantum} is roughly $O(\sqrt{n}\text{polylog}(n))$.
Thus the cost of Eq.~(\ref{eq:total_cost}) is:
\begin{align}
	O(\log(1/\epsilon)) \times (O(n)+O(\sqrt{n}\text{polylog}(n))+ O(n^2)) \nonumber \\
	= O(n^2\log(1/\epsilon))	. \label{eq:QMRMQLS_cost_dense}
\end{align}

\subsection{QMRM-QLS $\II$} \label{subsec:IIMHHL2}
In QMRM-QLS $\I$, the elements of the solution $\tilde{d}_m$ are all non-negative because of $Z$-basis measurements.
Using $\tilde{d}_m$ may cause an unstable accuracy improvement as shown in Fig.~\ref{fig:A2x=b2}.
We use QMRM-QLS $\II$ to stabilize accuracy improvement using $Z$-basis measurements.
Here, QMRM-QLS $\II$ is shown in Algorithm \ref{algo:IIMHHLver2}.
The idea to stabilize the accuracy improvement is as follows: shift the residual $r_{m+1}$ so that the elements of the next solution state $\ket{d_{m+1}}$ are non-negative.
Using a shift vector $x_{\text{shift}, m}$ in the $m$-th iteration, $f_m$ and $g_m$ are defined as follows: 
\begin{eqnarray}
	f_m(x_m, x_{\text{shift}, m}) := b - A x_m + A x_{\text{shift}, m}, \label{eq:IIMHHLv2:f} \\
	g_m(\tilde{d}_m, x_{\text{shift}, m}) := c_1 e^{ic_2}\tilde{d}_m - x_{\text{shift}, m}. \label{eq:IIMHHLv2:g}
\end{eqnarray}

If the shift vector $x_{\text{shift}, m}$ is the zero vector for all steps, QMRM-QLS $\II$ is the same as QMRM-QLS $\I$.
The elements of the shift vector are required to be all non-negative.
We consider these magnitudes to be close to the update values $g_m(\tilde{d}_m, x_{\text{shift}, m})$.
Thus, one possible shift vector can empirically be defined as follows:
\begin{equation} \label{eq:shift}
	x_{\text{shift}, m+1} = 
	\frac{\norm{g_m(\tilde{d}_m, x_{\text{shift}, m})}}
	     {\norm{g_{m-1}(\tilde{d}_{m-1}, x_{\text{shift}, m-1})}}
	\abs{g_m(\tilde{d}_m, x_{\text{shift}, m})},
\end{equation}
where $\abs{ \cdot }$ takes absolute values for each element.
This shift vector can reduce the negative values of the elements of the solution.
Note that $g_0(\tilde{d}_0, x_{\text{shift}, 0})$ is used instead of $g_{-1}(\tilde{d}_{-1}, x_{\text{shift}, -1})$ at $m=0$ in the experiments in Sec \ref{sec:experiments}.

The calculation costs of $f_m$ and $g_m$, respectively, Eqs. (\ref{eq:IIMHHLv2:f}) and (\ref{eq:IIMHHLv2:g}), are still $O(n)$ for a sparse matrix and $O(n^2)$ for a dense matrix.
Therefore, the cost of QMRM-QLS $\II$ for a sparse matrix and a dense matrix are, respectively, the same as Eqs~(\ref{eq:QMRMQLS_cost_sparse}) and (\ref{eq:QMRMQLS_cost_dense}).

From Eqs.~(\ref{eq:convergence_condition_of_x}) and (\ref{eq:xm+1}), the current solution $x_m$ can be refined when $\norm{d_m - g_m(\tilde{d}_m, x_{\text{shift}, m})} < 1$ which is the same as the QMRM-QLS $\I$.
\begin{algorithm}[H]
	\caption{QMRM-QLS $\II$}
	\label{algo:IIMHHLver2}
	\begin{algorithmic}[1]
	
		\Require{matrix $A$, vector $b$, target accuracy $\epsilon$, the number of measurements $N_{\text{shots}}$, running accuracy $\tilde{\epsilon}$}
		\Initialize{
			$x = 0$ // initial guess

			$m = 0$ // iteration count

			$x_{\text{shift}, 0} = 0$ // initial shift
		}
		
		\While{True}
			\State (a) $r_m = f_m(x_m, x_{\text{shift}, m})$
			
			\If{$\norm{r_m} < \epsilon$}
				\State break
			\EndIf		
		
			\State (b) {\it approximation $\tilde{d}_m$ by running $\mathcal{QLS}(\tilde{\epsilon}, A, \ket{r_m})$ and measuring $N_{\text{shots}}$}
			
			\State (c) $x_{m+1} = x_m + g_m(\tilde{d}_m, x_{\text{shift}, m})$
		
			\State Determine next $x_{\text{shift}, m+1}$
		
			\State $m \leftarrow m + 1$
		
		\EndWhile

		\Ensure{solution $x$}	
	
		\end{algorithmic}
\end{algorithm}

\section{Numerical Experiments} \label{sec:experiments}
In this section, we show some numerical experiments of QMRM-QLS $\I$ and $\II$.
The HHL algorithm is used as a quantum linear solver engine in our numerical experiments.

\subsection{Problems settings}
Two invertible $4 \times 4$ Hermitian matrices, and the solution vector $x$ are, respectively, as follows:
\begin{align}
	A_1 &= I \otimes I + 2 Z \otimes Z + 0.2 X \otimes X, \\
	A_2 &= Z \otimes Z + 0.5 X \otimes H, \text{and}\\
	x   &= 
	\begin{bmatrix}
		-10 & 1 & 0.1 & 0.01	
	\end{bmatrix}^T,
\end{align}
where $I$ is the identity, $X$ and $Z$ are Pauli matrices, and $H$ is the Hadamard matrix.
The condition numbers of $A_1$ and $A_2$ are, respectively, $4.000$ and $1.899$.
First, $b_1$ and $b_2$ are, respectively, prepared by calculating $b_1 = A_1 x$ and $b_2 = A_2 x$.
Next, the two linear systems, the systems $\I: A_1 x = b_1$ and $\II: A_2 x = b_2$, are solved, respectively.
We evaluate the solution using relative errors as follows:
\begin{align}
	\frac{\norm{x-x_{j,m}}}{\norm{x}}, (j=1, 2),
\end{align}
where $x_{1, m}$ and $x_{2, m}$ are, respectively, the $m$-th solution of the system $\I$ and $\II$ in QMRM-QLS.
Three tasks for performance analysis are listed as follows:
\begin{itemize}
	\item[(1)] To investigate the ideal performance of the QMRM-QLS $\I$, the solution state $\ket{d_m}$ in Eq.~(\ref{eq:x_HHL}) is used instead of the solution $\tilde{d}_m$;
	
	\item[(2)] To investigate the statistical behavior of QMRM-QLS $\I$ and $\II$, 10 calculations of 100 iterations are performed with $N_{\text{shots}} = 10^{4}$ measurements, and 6 qubits for the working register in the phase estimation are used; and

	\item[(3)] To compare the total number of measurements $N_{\text{total}}$ needed for QMRM-QLS and the HHL, the HHL is performed with 14 qubits for the phase estimation in order to prepare a solution with sufficient accuracy.
\end{itemize}

Note that the number of measurements $N_{\text{shots}}$ includes both the number of the HHL successes and failures.
Hence the total number of measurements $N_{\text{total}}$ also includes both.
For simulations of the HHL algorithm without noise, we use qiskit which is open-source software for simulating quantum computing \cite{qiskit_short}.
For all numerical experiments, we use the double-precision floating-point number.
Therefore, there is an upper limit to the number of significant digits, which is approximately $16$ digits in decimals \cite{goldberg1991What}.

\subsection{Results \& Discussions}
\begin{figure}
	\centering
	\includegraphics[scale=0.30]{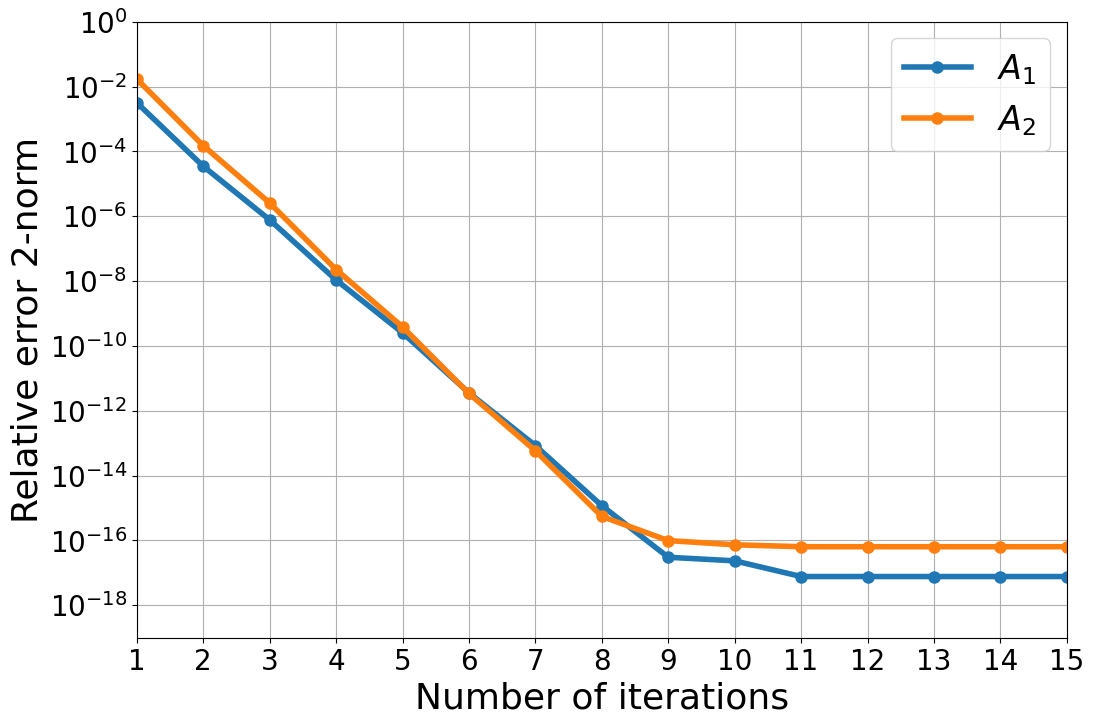}
	\caption{The iteration history of relative errors solving the system $\I$ ($A_1$, blue) and $\II$ ($A_2$, orange), respectively, by using QMRM-QLS $\I$. The solution state $\ket{d_m}$ is used instead of $\tilde{d}_m$.}
	\label{fig:statevector}
\end{figure}

Figure \ref{fig:statevector} shows relative errors of the solution $x$ of the system $\I$ and $\II$, respectively, using QMRM-QLS $\I$.
We use the solution state $\ket{d_m}$ instead of measurement results $\tilde{d}_m$ in the iterative processes.
The horizontal axis is the number of iterations and the vertical axis is the relative error.
The first iterative results indicate the results of HHL itself, and they are approximately 2 digits accurate.
Namely, 6 qubits are enough for the phase estimation to obtain a solution with 2 digit accuracy in our problem settings.
Here, QLS-QLS $\I$ obtains $17$ and $16$ digits accuracy in total, improving the accuracy by approximately $2$ digits per iteration.
The reason for the saturation of accuracy from the 8th iteration is due to the double-precision floating-point number used in these calculations.
The linear decrease in the relative errors is the expected behavior of QLS-QLS $\I$.

\begin{figure}
	\centering
	\includegraphics[scale=0.30]{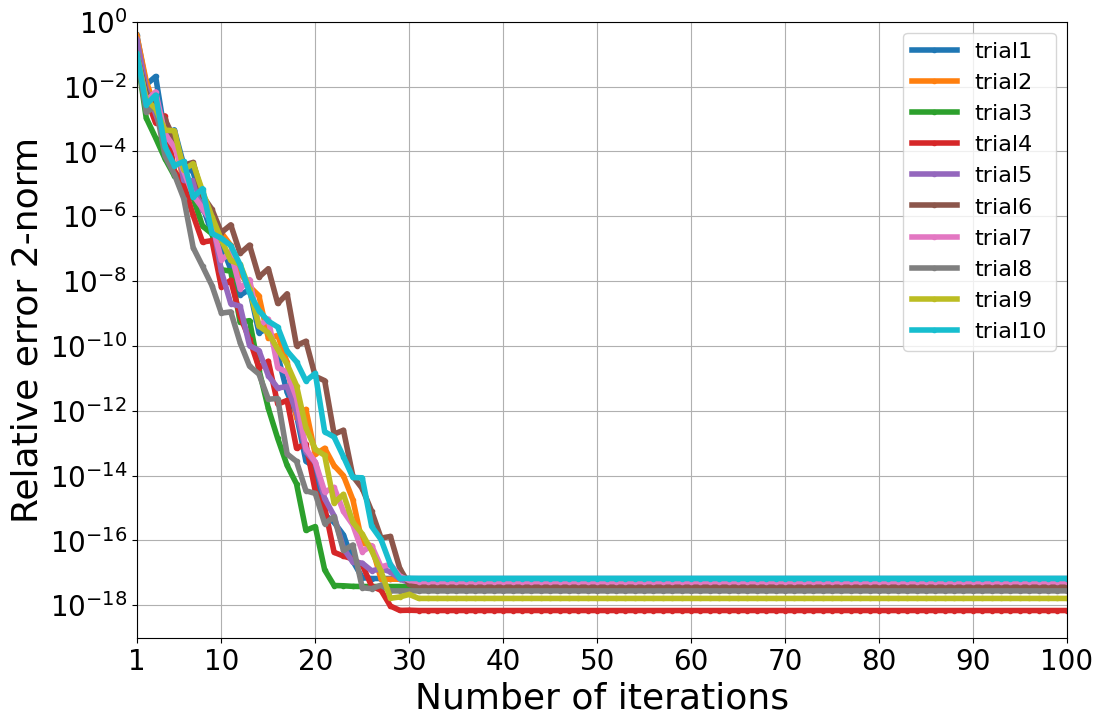}
	\caption{The iteration history of relative errors solving the system $\I: A_1 x = b_1$ using QMRM-QLS $\I$. Here, the same 10 calculations (trial $1 \sim 10$) are performed.}
	\label{fig:A1x=b1}
\end{figure}

\begin{figure}
	\centering
	\includegraphics[scale=0.30]{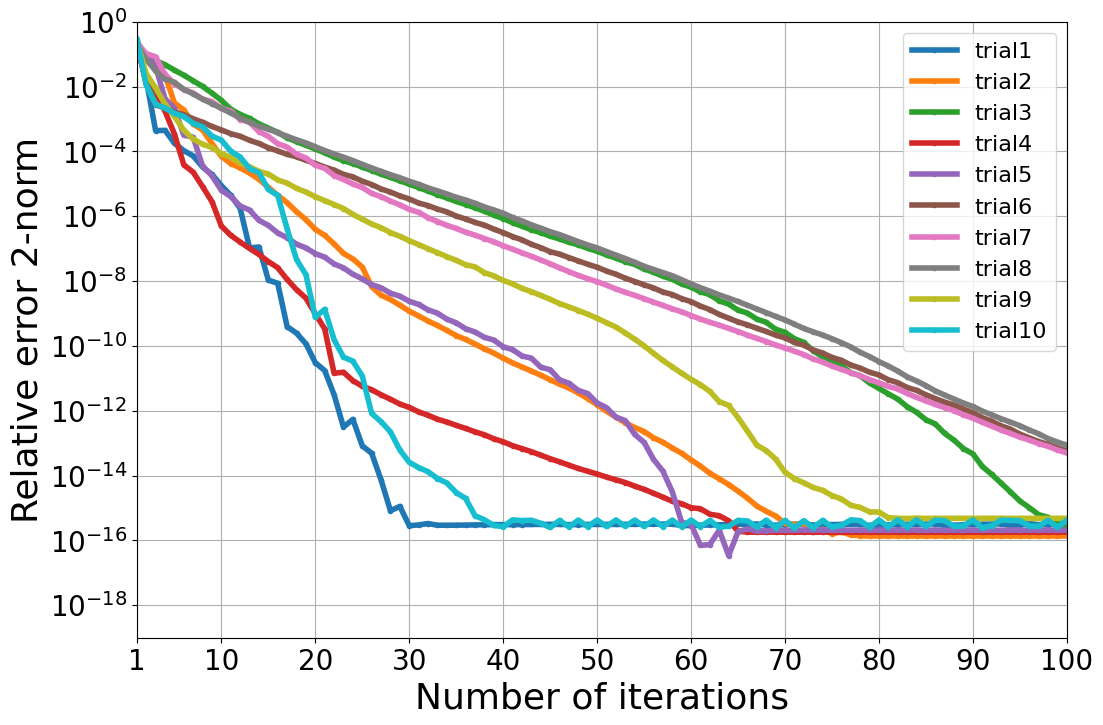}
	\caption{The iteration history of relative errors solving the system $\II: A_2 x = b_2$ using QMRM-QLS $\I$. Here, the same 10 calculations (trial $1 \sim 10$) are performed.}
	\label{fig:A2x=b2}
\end{figure}

Figures \ref{fig:A1x=b1} and \ref{fig:A2x=b2} show relative errors of solutions $x$ solving, respectively, the systems $\I$ and $\II$ using QMRM-QLS $\I$.
Both systems are solved 10 times to investigate the statistical behavior.
The differences in the 10 calculation results occur due to the measurements. 
For all 10 calculations in Fig.~\ref{fig:A1x=b1}, the errors converge by the 30th iteration, and about 17 or 18 digit accurate solutions are obtained.
On the other hand, in Fig.~\ref{fig:A2x=b2}, the errors decrease, and at least 13 digit accurate solutions are obtained.
The convergence speeds vary compared with the results in Fig.~\ref{fig:A1x=b1}.
This indicates that more iterations are needed for high-accuracy solutions, which also means an increase in the number of measurements.
However, the measurement cost is still $O(n\log(1/\epsilon))$.

\begin{figure}
	\centering
	\includegraphics[scale=0.30]{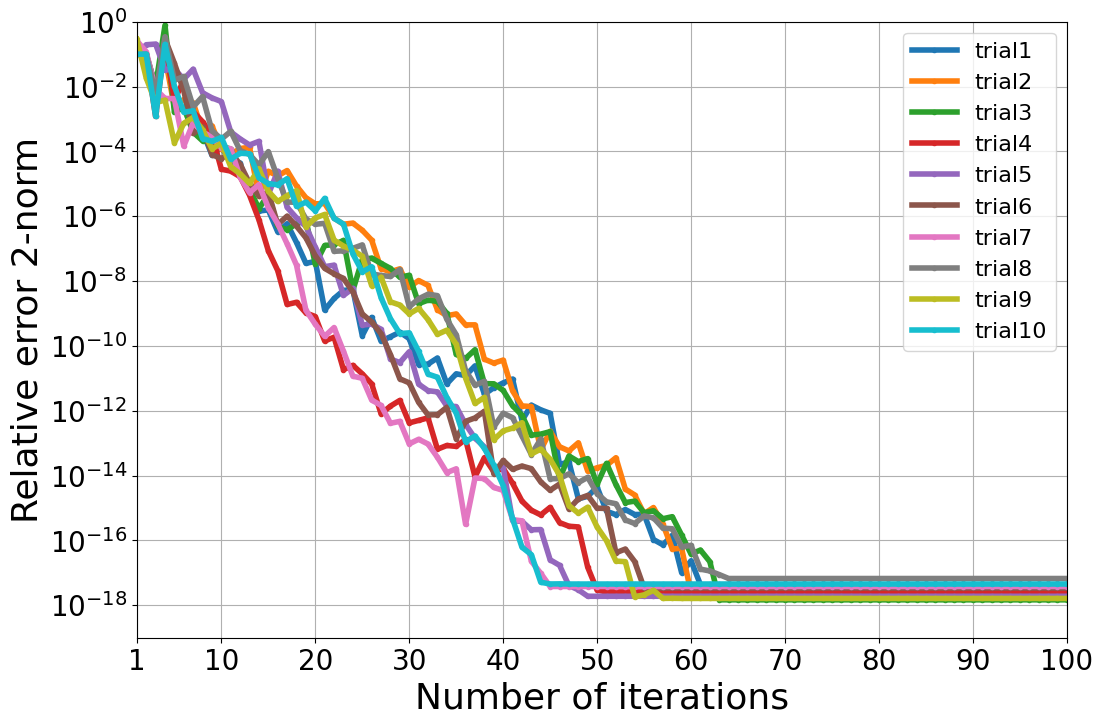}	
	\caption{The iteration history of relative errors solving the system $\I: A_1 x = b_1$ using QMRM-QLS $\II$. Here, the same 10 calculations (trial $1 \sim 10$) are performed.}
	\label{fig:A1x=b1:shift}
\end{figure}

\begin{figure}
	\centering
	\includegraphics[scale=0.30]{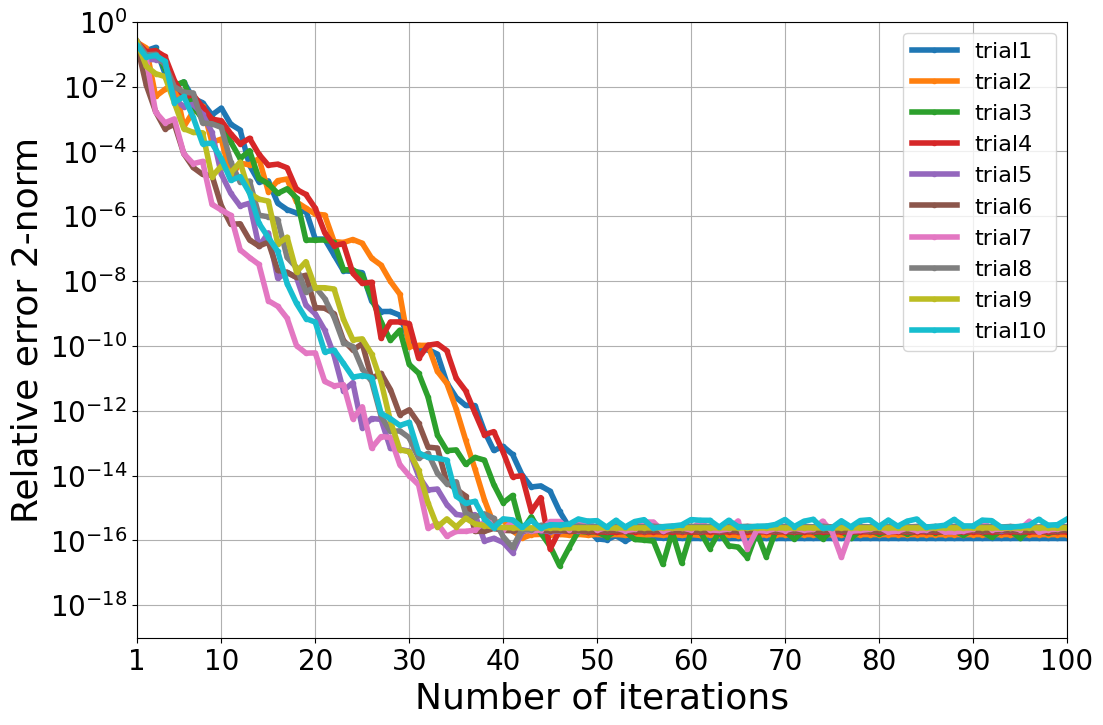}	
	\caption{The iteration history of relative errors solving the system $\II: A_2 x = b_2$ using QMRM-QLS $\II$. Here, the same 10 calculations (trial $1 \sim 10$) are performed.}
	\label{fig:A2x=b2:shift}
\end{figure}

Figures \ref{fig:A1x=b1:shift} and \ref{fig:A2x=b2:shift} show relative errors of the solutions $x$, respectively, solving the systems $\I$ and $\II$ using QMRM-QLS $\II$.
For both results, 16 digit accurate solutions are obtained by about 50th to 60th iterations, respectively.
In Fig.~\ref{fig:A1x=b1:shift}, the error convergence speeds are slightly slower than those in Fig.~\ref{fig:A1x=b1}.
On the other hand, the convergence speeds shown in Fig.~\ref{fig:A2x=b2:shift} are stable compared with those without the shift shown in Fig.~\ref{fig:A2x=b2}.
Therefore, the shift operations in QMRM-QLS $\II$ are expected to stabilize the convergence speed of the errors while using $Z$-basis measurements.
However, more iterations are needed.

Regarding the unstable error convergence, as shown in Fig.~\ref{fig:A2x=b2}, it is difficult to determine what matrix properties such as the condition number, the number of non-zeros, etc,  cause the unstable convergence.
However, we consider that those unstable convergence behaviors may be caused by little differences between the exact $d_m$ and $g_m(\tilde{d}_m)$ to refine the solution.
By taking the shift operation using functions $f$ and $g$, we expect the stability of the error convergence speed to be improved.
We also expect that the appropriate choice of $f$ and $g$ can stabilize the error convergence speed.

\begin{figure}
	\centering
	\includegraphics[scale=0.30]{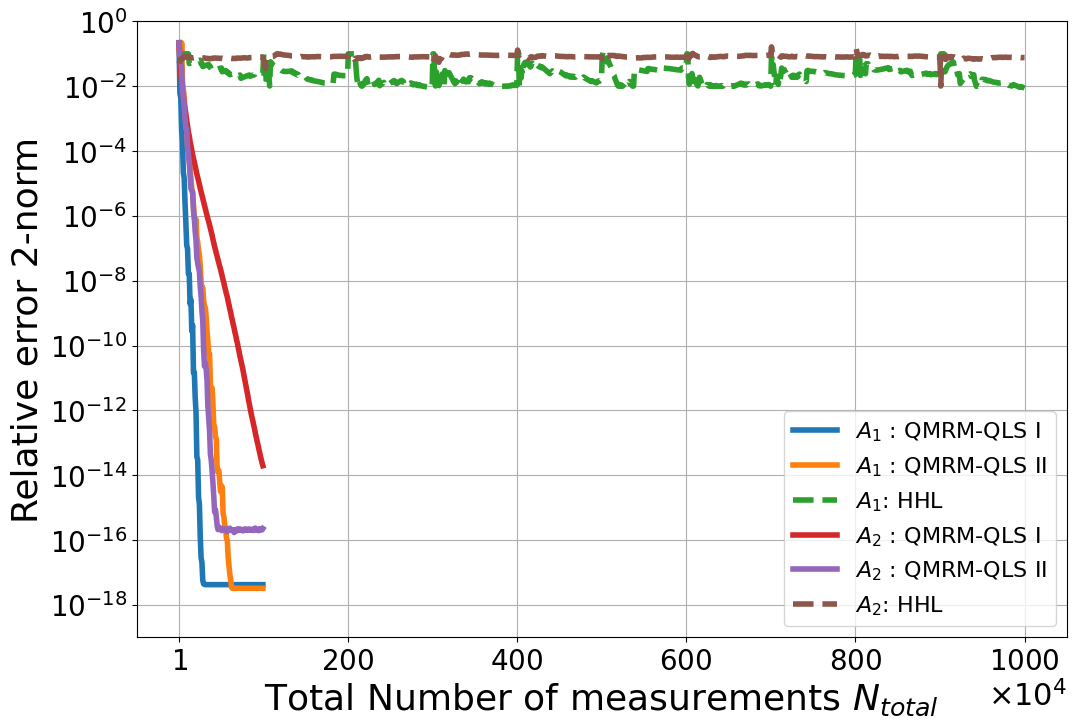}
	\caption{The relationship between the relative errors and the number of measurements up to $1000\times 10^4$ measurements using QMRM $\I, \II$, and HHL solving the systems $\I$ and $\II$.}
	\label{fig:shots_1000}
\end{figure}

Figure~\ref{fig:shots_1000} shows the relationship between the total number of measurements $N_{\text{total}}$ and the relative errors.
The horizontal axis is the total number of measurements $N_{\text{total}}$ and the vertical axis is the relative errors.
The solid lines are, respectively, the average of the results shown in Figs~\ref{fig:A1x=b1}--\ref{fig:A2x=b2:shift} up to the 100th iteration.
The dashed lines are, respectively, the results solving the system $\I$ and $\II$ using the HHL with 14 qubits to eliminate any effect on the accuracy except the number of measurements.
Here, $1000\times 10^4$ measurements in Fig.~\ref{fig:shots_1000} are performed in HHL calculations.
We numerically confirmed that the HHL can obtain a $7$ digit accurate solution $x$ solving the system $\I$ and a $6$ digit accurate solution $x$ solving the system $\II$, respectively, using 14 qubits for the phase estimation and solution state $\ket{x}$ in Eq.~(\ref{eq:x_HHL}).
The reason the HHL accuracy does not improve is that the number of measurements is not enough even with $1000 \times 10^4$ measurements.
This indicates that a high-accuracy solution is hard to obtain using only the HHL even with a high-accuracy quantum state.
On the other hand, QMRM-QLS can achieve a more accurate solution with a smaller number of measurements than that of the HHL.
In addition, QMRM-QLS obtains high-accuracy results using only $6$ qubits for the phase estimation.
This means that quantum resources to run HHL are reduced.

\section{Conclusions} \label{sec:conclusion}
In this paper, we propose quantum multi-resolution measurement (QMRM) to reduce the number of measurements required for accuracy.
We show the number of measurements for the solution with an accuracy $\epsilon$ in QMRM is $O(n\log(1/\epsilon))$ in Theorem~\ref{theorem:measurements}.
From Eq.~(\ref{eq:accuracy_relation_v2}), we show that QMRM can reduce quantum resources for the running accuracy of a quantum algorithm.
We propose quantum linear solvers QMRM-QLS $\I$ and $\II$ using the generator $f$ and the adjustor $g$ with QMRM, and we conduct some numerical experiments using them.
These numerical results show that we obtain about a $16$ digit accurate solution in both QMRM-QLS $\I$ and $\II$ using $6$ qubits for the quantum phase estimation in the HHL algorithm and $N_{\text{total}} = 10^6$ measurements.
These results indicate that the QMRM-QLS can obtain a more accurate solution with a smaller number of measurements and smaller quantum resources than that of the HHL algorithm itself.
Moreover, we find that it is useful for stable calculations to take the shift operation in QMRM-QLS $\II$.
It is interesting to find a more effective generator $f$ and adjustor $g$ to stabilize and accelerate the solver.
Using the effective $f$ and $g$ can reduce the number of measurements.

In addition, QMRM can separate a quantum circuit into a fixed part and a variable part.
Reusing the fixed part can save the construction time of quantum circuits in iterations.
On the other hand, QMRM must reconstruct the variable part, hence QMRM needs its reconstruction time at every iteration.
We expect that QMRM can also reduce the total computation time including the (re)construction time of quantum circuits.

For the problem size $n$, the cost of solving a linear system of equations is $O(n)$ even in classical calculations.
Based on the many applications of a linear system and the refining of the solution, we believe that QMRM can be a useful tool to obtain high-accuracy solutions and expensive numerical computations.

\section*{Acknowledgements}
The authors would like to express their sincere gratitude to Jason Ginsburg for valuable comments, which improve the presentation of this manuscript.

\bibliography{ref}

\end{document}